\documentclass[12pt,preprint]{aastex}
\newcommand{\msun}{M_{\odot}}

\renewcommand{\S}{Section}

\begin{document}
\slugcomment{{\scriptsize Accepted for the ApJS Special Issue on the Chandra Carina Complex Project}}

\title{A {\it Chandra} ACIS Study of the Young Star Cluster Trumpler 15 in Carina and Correlation with Near-infrared Sources}

\author{Junfeng Wang,\altaffilmark{1} Eric
  D. Feigelson,\altaffilmark{2} Leisa K. Townsley,\altaffilmark{2}
  Patrick S. Broos,\altaffilmark{2} Konstantin
  V. Getman,\altaffilmark{2} Scott J. Wolk,\altaffilmark{1} Thomas
  Preibisch,\altaffilmark{3} Keivan G. Stassun,\altaffilmark{4,5}
  Anthony F. J. Moffat,\altaffilmark{6} Gordon Garmire,\altaffilmark{2}
 Robert R. King,\altaffilmark{7} 
  Mark~J. McCaughrean,\altaffilmark{7,8} and Hans
  Zinnecker\altaffilmark{9}}

\altaffiltext{1}{Harvard-Smithsonian Center for Astrophysics, 60
Garden St, Cambridge, MA 02138}

\altaffiltext{2}{Department of Astronomy \& Astrophysics, The
Pennsylvania State University, 525 Davey Lab, University Park, PA
16802}

\altaffiltext{3}{Universit\"ats-Sternwarte M\"unchen,
  Ludwig-Maximilians-Universit\"at, Scheinerstr.~1, 81679 M\"unchen,
  Germany}

\altaffiltext{4}{Department of Physics \& Astronomy, Vanderbilt University, Nashville, TN 37235}

\altaffiltext{5}{Department of Physics, Fisk University, 1000 17th Ave. N., Nashville, TN 37208}

\altaffiltext{6}{D\'{e}partement de Physique, Universit\'{e} de Montr\'{e}al, Succursale Centre-Ville, Montr\'{e}al, QC, H3C 3J7, Canada}

\altaffiltext{7}{Astrophysics Group, College of Engineering, Mathematics, and Physical Sciences, University of Exeter, Exeter EX4 4QL, UK}

\altaffiltext{8}{ESA Research and Scientific Support Department,
  ESTEC, Postbus 299, 2200 AG Noordwijk, The Netherlands}

\altaffiltext{9}{Astrophysikalisches Institut Potsdam, An der
            Sternwarte 16, 14482 Potsdam, Germany}

\begin{abstract}

Using the highest-resolution X-ray observation of the Trumpler 15 star
cluster taken by the {\em Chandra X-ray Observatory}, we estimate the
total size of its stellar population by comparing the X-ray luminosity
function of the detected sources to a calibrator cluster, and identify
for the first time a significant fraction ($\sim$14\%) of its
individual members.  The highest-resolution near-IR observation of
Trumpler 15 (taken by the HAWK-I instrument on the VLT) was found to
detect most of our X-ray selected sample of cluster members, with a
$K$-excess disk frequency of $3.8\pm 0.7\%$.  The near-IR data, X-ray
luminosity function, and published spectral types of the brightest
members support a cluster age estimate (5--10 Myr) that is older than
those for the nearby Trumpler 14 and Trumpler 16 clusters, and suggest
that high-mass members may have already exploded as supernovae.  The
morphology of the inner ${\sim}$0.7 pc core of the cluster is found to
be spherical.  However, the outer regions (beyond ${\sim}$2 pc) are
elongated, forming an ``envelope'' of stars that, in projection,
appears to connect Trumpler 15 to Trumpler 14; this morphology
supports the view that these clusters are physically associated.
Clear evidence of mass segregation is seen.  This study appears in a
\anchor{TBD}{Special Issue} of the \apjs\/ devoted to the {\em
  Chandra} Carina Complex Project (CCCP), a 1.42 square degree {\em
  Chandra} X-ray survey of the Great Nebula in Carina.

\end{abstract}


%=============================================================================
\section{Introduction}

Trumpler 15(=C 1042-591 = Lund 558 = Ocl 825) is a compact and rich
young open cluster located near the northeast edge of the Carina
Nebula (see Smith \& Brooks 2008 for a review).  In contrast to other
young clusters in the region, such as Trumpler 14 and Trumpler 16, it
is poorly studied: only a few optical and near infrared (near-IR)
investigations exist in the literature and $\sim$20 candidate members
have spectral classification \citep{M88,Skiff09}, limited to the
brightest OB stars.

The optical and near-IR studies are mainly hampered by the heavy
contamination from background stars, since there is little dust in the
region and no dense molecular cloud behind the cluster \citep{C02}.
The basic properties of the cluster from major studies in the
literature since the 1960s are summarized in Table~\ref{tbl:basic}.
There was considerable disagreement on the distance
\citep[e.g.,][]{T71,W73} and age of Trumpler 15 until recent studies
with modern CCD photometry \citep[e.g.,][]{C02,T03}.  Consequently,
the relationship of Trumpler 15 to the other clusters (e.g., Trumpler
14/16) located closer to the luminous blue variable (LBV) star $\eta$
Carinae \citep{DH97} has been unclear.  There is strong observational
evidence that massive star formation in the Carina Nebula has been
active for several million years and is ongoing (Vazquez et al. 1996,
Smith et al. 2000, Ascenso et al. 2007).  For Trumpler 15, Dias et
al. (2002) list a cluster age of 8 Myr, while Tapia et al. (2003)
suggest a range of ages between 4 Myr and 30 Myr with a median value
of $\sim$8 Myr.  This is consistent with a picture where Trumpler 15
likely resulted from an early burst of star formation in the complex.

Perhaps due to the historically sparse coverage of optical and near-IR
studies, there has been no previous X-ray study of this star cluster.
High-resolution X-ray images of young stellar clusters are effective
in detecting large populations of pre-main sequence (PMS) stars with
little contamination from Galactic field stars (see Feigelson et
al. 2007 for a review).  X-ray selection is particularly useful
because it traces magnetic flaring that is largely insensitive to the
presence or absence of a protoplanetary disk.  X-ray samples are thus
complementary to infrared-excess samples which locate young stars with
dusty disks.

In this paper we present a {\em Chandra} imaging study of Trumpler 15,
aiming to obtain the first X-ray census of its member stars.  \S~2
gives a brief description of our {\em Chandra} observation, and
presents a list of X-ray selected young stellar members.  Using X-ray,
optical, and near-IR data we study in \S~3 the cluster morphology,
X-ray luminosity function (XLF), extinction, and age. \S~4 presents
notes on some interesting stars. Our findings are summarized in \S~5.
As in all CCCP papers \citep{T11a}, we adopt a distance of 2.3~kpc to
Trumpler 15 (Smith 2006) throughout the paper; thus, 1\arcmin\/ on the
sky corresponds to 0.7~pc.

\section{Observations and Data Analysis}\label{sec:obs.sec}

Trumpler 15 was observed as part of the {\em Chandra Carina Complex
  Project} (CCCP; PI: L. Townsley), a large mosaic of {\em Chandra}
Advanced CCD Imaging Spectrometer I-array (ACIS-I)
observations. \citet{T11a} provide a complete overview of the CCCP.
One CCCP pointing (ObsID 9484, a 60 ks integration obtained on August
19, 2008) is centered on Trumpler 15; five other pointings cover the
edges of the cluster (Figure 1a).

\citet{Broos11a} describe the custom data preparation, source
detection, and source extraction (14369 ACIS point sources)
procedures\footnote { Most of the software required to produce the
  CCCP source catalog is publicly available in the {\it ACIS Extract}
  package
  (\url{http://www.astro.psu.edu/xray/acis/acis\_analysis.html})
  \citep{Broos10}.  }  applied to the CCCP data; \citet{Broos11b}
present a source classifier for Carina X-ray source membership that
calculates probabilities that a given X-ray source is a foreground
star, a background star, an extragalactic source, or a Carina member
with careful evaluation of the contaminating populations
\citep{Getman11}.  We adopt these classifications in this study,
recognizing that they are not perfect; some false positives
(contaminants incorrectly identified as Carina members) and false
negatives (true Carina members unclassified or incorrectly identified
as contaminants) are present.  X-ray point source fluxes and
absorptions have been estimated with a non-parametric method using
median X-ray energy and observed photon flux (XPHOT; Getman et
al. 2010).

\citet{F10} investigated the spatial distribution of the X-ray sources
classified as likely Carina members and the clustering structure based
on maps of smoothed source surface density.  Trumpler 15 is one of the
20 identified principal clusters (CCCP-Cl\#8; Figure~1b in Feigelson
et al. 2011) obtained by smoothing a spatially complete sample of
likely Carina members with a Gaussian kernel ($\sigma=30$\arcsec\/,
corresponding to $FWHM=0.8$~pc).  In this paper, we adopted a
footprint for the Trumpler 15 cluster that is slightly larger than
region CCCP-Cl\#8.  The footprint traces the lowest source density
contour presented by \citet[][Figure~1]{F10} and a boundary
drawn by-eye to separate Trumpler 15 from Trumpler 14 (see Figure 11
in Townsley et al. 2011a for a global view of Carina's
historically-recognized constituent clusters).

The adopted Trumpler 15 region contains 841 CCCP sources, listed in
Table~2.  Detailed source properties can be extracted from the CCCP
catalog \citep{Broos11a}.  The stellar sample we discuss henceforth
consists of the 829 sources (98.6\%) classified as probable Carina
members \citep{Broos11b}, identified by the ``H2'' class in Table~2.
\citet{Broos11b} estimate that a few percent of the ``H2'' sources in
the rich cluster region may in fact be foreground Galactic field stars
or background extragalactic sources.

Counterparts to CCCP sources have been identified in many optical and
IR studies \citep{Broos11a}.  The near-IR data used in this study
(\S~\ref{sec:NIR}) were obtained from the deep HAWK-I\footnote{The
  High Accuity Wide-field K-band Imager, an instrument on the ESO 8~m
  Very Large Telescope (VLT). See \citet{KP08} for more information on
  HAWK-I.}  survey \citep{P10}. A total of 748 (89\%) of the ACIS
sources have near-IR counterparts with valid $JHKs$ photometry in at
least one band; 738 (88\%) have valid $JHKs$ photometry in all three
bands.  Details of the HAWK-I photometric quality is described in
\citet{P10} (and references therein); briefly, most sources with
$J\lesssim 21.2$, $H\lesssim 20.3$, and $Ks\lesssim 19.3$ have formal
photometric uncertainties $<0.1$ mag ($S/N>10$ detections).

\section{X-ray Cluster Properties}

\subsection{Cluster Morphology}

Figures 1 and 2 show the cluster region in three bands.  Figure 1a is
a wide-area view of the 2MASS survey with the CCCP ACIS fields
delineated.  The two brightest stars in the field are red supergiants
(RSGs), RT~Car (M2Iab) and BO~Car (M4Iab), at $K \sim 1.5$ mag; no
firm conclusion has been drawn about their membership to Trumpler 15
\citep{T03,S06}.  The Trumpler 15 cluster lies in the middle of the
figure, identified as a sparse concentration of a dozen 2MASS stars.
A much richer cluster is seen in the deeper HAWK-I image \citep{P10}.
The cluster is comparably populous in the CCCP X-ray image, shown in
Figure 1b, analyzed here.

Figure 2 shows a smoothed contour map of the X-ray sources on the
optical Digital Sky Survey (DSS) image.  \citet{C02} analyzes deep
$UBV$ images and finds that the cluster is compact, with a core radius
of $\sim$2\arcmin\/ (1.3~pc).  \citet{T03} measured a cluster size of
$r=5.3$\arcmin\/ using $V$-band star counts.  However, contamination
by foreground and background stars prevents study of the low density
outer regions of the cluster, which are readily seen in the X-ray
band. The X-ray cluster also appears compact and spherical in the
inner $r=1\arcmin$ core, but becomes elongated at larger distances
($\sim 6\arcmin$ along the north-south direction vs. $\sim 4\arcmin$
east-west), as outlined by the contours of source density (Figure 2)
and as seen in the X-ray image (Figure 1b).  Beyond $r=3\arcmin$,
there is an ``envelope'' of more dispersed stars extending south-west
towards Trumpler 14.  The continuous distribution of stars between
Trumpler 14 and 15 \citep{F10,T11a} strongly indicates that Trumpler
15 indeed is part of the Carina complex rather than the spatial
superposition of a more distant cluster (Walborn 1995).  Both Trumpler
14 and 15 are dominated by a central core where the brightest massive
stars are concentrated, whereas Trumpler 16 is comprised of a number
of subclusters with widely dispersed massive stars \citep{Wolk11}.  We
also note that Trumpler 14 has a much higher star density in its core
than either Trumpler 15 or 16.

The X-ray defined cluster center of Trumpler 15
($\alpha,\delta$)$_{\rm J2000}$=(10$^h$44$^m$43.$^s$8,
-59$^{\circ}$21$^{\prime}$42$^{\prime\prime}$) (Feigelson et al. 2011)
is in good agreement with the cluster center measured by optical
studies ($\alpha,\delta$)$_{\rm J2000}$=(10$^h$44$^m$43.$^s$2,
-59$^{\circ}$21$^{\prime}$49$^{\prime\prime}$) \citep[e.g.,][]{C02},
considering that the optical position is predominantly determined by
the locations of bright OB stars. The radial density profile of the
X-ray cluster is shown in Figure~3, derived by counting stars in
concentric annuli around the X-ray cluster peak density.  For
comparison, the optical radial profile in \citet{C02} is also shown.
Both are sharply peaked in the inner $r<1\arcmin$ region, but the
X-ray source density is 4 times higher than that in the optical.  This
is mainly due to the effectiveness of the X-ray survey locating the
low mass PMS cluster members, many of which are missed in the optical
due to crowding and bright sources in the core.  While the density
profile falls smoothly to 2 stars per arcmin$^2$ in the optical, the
X-ray profile extends to 0.5 stars per arcmin$^2$ and shows a bump at
$r=1.5\arcmin$.  This density enhancement likely arises from a clump
of stars to the north of the cluster center (indicated with a circle
in Figure~\ref{fig:himass}a), causing an elongation of the X-ray
stellar density contours towards the north. This clump of stars is not
apparent in the optical DSS image (Figure~2).

Figure~\ref{fig:himass}a shows the spatial distribution of the O and B
stars (ranging from B9 to O9 in Trumpler 15) from the Catalogue of
Stellar Spectral Classifications \citep{Skiff09}.  The X-ray detected
massive stars will be further discussed in \S~\ref{sec:individual}.
Figure~\ref{fig:himass}a also presents the spatial distribution of the
X-ray selected K-band excess stars (\S~\ref{sec:NIR}) that retain
optically thick circumstellar disks.  These represent the younger
stellar population.  They are not concentrated towards the cluster
center, but follow the broad envelope of the Trumpler 15 X-ray source
distribution, including the extension southward towards Trumpler 14.

Cumulative radial distributions for the massive stars and the CCCP
sample with X-ray-detected massive stars removed, shown in
Figure~\ref{fig:himass}b, reveal clear mass segregation.  For example,
the enclosed fractions of massive and CCCP stars are 43\% and 17\% at
$r = 1\arcmin$, and are 78\% and 46\% at $r = 2\arcmin$.  The
probability that the two populations actually have the same radial
profile was found to be only 0.4\%, via a 2-sample Kolmogorov-Smirnov
statistic.

\subsection{X-ray Luminosity Function}\label{sec:xlf}

Although the massive stars in Trumpler 15 were identified by early
studies \citep[e.g.,][]{G68}, investigation of the low-mass population
(less massive than 4~$\msun$) has been limited by overwhelming
contamination from background stars \citep{T03}, thus the total
stellar population in Trumpler 15 is still largely unknown.
\citet{Feigelson05} suggested that the measured XLF in a young stellar
cluster can be viewed as the convolution of the Initial Mass Function
(IMF) and the mass--luminosity correlation (which is measured in the
Chandra Orion Ultradeep Project [COUP] studies; Getman et al. 2005,
Preibisch et al.\ 2005).  Using the best-studied Orion Nebula Cluster
XLF (COUP XLF) as a calibrator, the XLF of a cluster can be used to
probe its IMF and to estimate the total X-ray emitting
population. Such a population analysis has been performed in several
other studies
\citep{Getman06,Wang07,Broos07,Wang08,Wang10,Winston10,Wolk10}. In a
similar XLF analysis performed here, we consider the total band
(0.5--8 keV) absorption corrected X-ray luminosity, $L_{t,c}$,
estimated by XPHOT (Getman et al.\ 2010; \S~2).

We consistently detect 15 count sources at any off-axis location in
the 60 ks exposure (see Figure 7 in Broos et al. 2011a), which allows
us to estimate the detection completeness limit in $L_{t,c}$. Assuming
an absorption column density $N_H=3\times 10^{21}$ cm$^{-2}$
\citep[corresponding to $A_V\sim 2$ mag, typical for Trumpler 15
  stars;][]{C02} and a thermal plasma with $kT = 1$ keV (typical for
young X-ray emitting low-mass stars; e.g., Preibisch et al. 2005), the
corresponding ACIS point source luminosity is $\log L_{t,c} \sim 30.4$
erg s$^{-1}$, calculated with the Portable Interactive Multi-Mission
Simulator (PIMMS\footnote{Available at
  \url{http://heasarc.nasa.gov/docs/software/tools/pimms.html}}) tool.

Figure~\ref{fig:XLF} shows the XLF for 316 Trumpler 15 sources for
which $L_{t,c}$ is available ($L_{t,c}$ cannot be obtained for many
faint sources) and membership in the Carina complex is likely (the
``H2'' classification of Broos et al. 2011a). Overplotted is a
template XLF obtained from the COUP unobscured population (839 cool
stars, excluding high-mass stars with spectral types earlier than B4;
Feigelson et al. 2005).  Seven X-ray detected O and B stars in
Trumpler 15 (\S~\ref{sec:individual}) were excluded to be consistent
with the COUP sample.

Comparing to the COUP XLF, the Trumpler 15 XLF turns over quickly at
$\log L_{t,c}<30.4$ due to incompleteness.  We model the XLF
($L_{t,c}$) with a power-law distribution, restricted to a range of
luminosities above this completeness limit.  Parameters of the
power-law are derived by directly fitting the luminosity values,
without binning, to the model as described by Maschberger \& Kroupa
(2009).  Our choice for the range of luminosities fit was informed by
the so-called ``stablilized P-P plot" produced by the fitting process
(Maschberger \& Kroupa 2009; Michael 1983).\footnote{The TARA package
  (\url{http://www.astro.psu.edu/xray/docs/TARA/}) provides a tool,
  {\em ml\_powerlaw}, that implements this bin-free fitting procedure
  and the stabilized P-P plot.}  The slope of the Trumpler 15 XLF
($\Gamma=-0.96\pm 0.13$) is largely consistent with that of the COUP
XLF ($\Gamma=-0.89 \pm 0.07$) in the luminosity range $30.4\lesssim
\log L_{t,c} \lesssim 31.0$ (erg s$^{-1}$).  The 1-$\sigma$ errors on
the power-law slopes were derived using the non-parametric bootstrap
method.  Matching the two cluster XLFs in the range of $30.4\lesssim
\log L_{t,c} \lesssim 31.0$ requires scaling up the ONC population by
a factor of $2.1 \pm 0.3$.  If we assume the ONC has $\sim 2800$ stars
(within a projected 2~pc radius, Hillenbrand \& Hartmann 1998), then
Trumpler 15 is inferred to have $\sim 5900$ stars. Note that this is
only an approximate measure of the cluster population, as quantitative
comparison between Trumpler 15 and the ONC depends on accurate
knowledge of their respective distances, physical extents, and
absorption characteristics.

However, if the more luminous bins ($\log L_{t,c} \gtrsim 31$) are
included in the fit, the slope of the Trumpler 15 XLF becomes much
steeper ($\Gamma=-1.27\pm 0.10$) than the COUP XLF
(Figure~\ref{fig:XLF}).  This cannot be an artifact of the detection
completeness limit.  The luminosity bins causing the steep slope in
the Trumpler 15 XLF are at the brighter end of the XLF, therefore
these sources are not missed due to sensitivity.  We further consider
the small differences between the X-ray detection efficiencies for
$K_s$-excess and non $K_s$-excess stars (Flaccomio et al 2003,
Preibisch et al 2005, Getman et al. 2009, Winston et al. 2010), and
for lightly and heavily obscured stars (Feigelson et al 2005).  We
have explored the XLFs with a smaller Trumpler 15 sample (307 stars)
that excludes the $K_s$-excess stars, and with another sample (262
stars) that excludes the sources with median photon energy $MedE\geq
2.0$ keV; both alternative samples show little impact on the slope of
the Trumpler 15 XLF when high luminosity bins are included
($\Gamma=-1.3\pm 0.1$).

Two mechanisms intrinsic to the Trumpler 15 cluster may be responsible
for the observed slope anomaly.  First, if Trumpler 15's formation
process was consistent with a standard IMF (e.g., Kroupa 2001), then
the steep slope may be related to its age.  The apparent absence of
stars earlier than O9 at the current epoch strongly suggests that
massive stars have been lost, presumably by evolving into supernovae
(SNe).  Specifically, a cluster with 5900 stars is expected to have an
average number of 11.5 stars with $M > 20 \msun$ assuming the Kroupa
(2001) IMF.  The probability of forming no such massive stars in
Trumpler 15 is $<10^{-5}$, based on $10^6$ random realizations of a
simulated cluster containing 5900 stars with masses distributed
according to the Kroupa (2001) IMF.  The notion that the cluster's
massive stars (which would be luminous X-ray sources) have been
destroyed is consistent with the cluster age estimate (5--10 Myr,
\S~3.3) and is supported by the presence of the O9III stars in the
cluster, which are possibly the post-main sequence members that will
soon evolve to supernovae explosions.  In addition, Povich et
al. (2011) found some evidence in the CCCP sample that intermediate
mass (2--10~$\msun$) stars are intrinsic X-ray emitters during their
PMS evolution, but then the emission diminishes in a few Myrs or less.
Since Trumpler 15 is significantly older than the ONC, our X-ray
sample likely misses some intermediate mass stars due to this
evolution, which can also contribute to a steeper slope in the
observed XLF.

Alternatively, the steep XLF slope may indicate that the Trumpler 15
IMF deviates from the Orion IMF with an excess of $\sim 120$ stars in
the luminosity range $30.4\lesssim \log L_{t,c} \lesssim 31.0$ erg
s$^{-1}$ or, equivalently, deficit of massive stars producing $\log
L_{t,c} \gtrsim 31.0$ erg s$^{-1}$.  Although we cannot exclude an
anomalous IMF, these extra stars are unlikely to be foreground
contaminants that are erroneously classified as Carina members because
essentially no stars are found close to the unreddened main sequence
locus in the NIR color-magnitude diagram of Trumpler 15 (see next
section).  To test whether the adjacent Trumpler 14 cluster may
contribute additional stars to Trumpler 15, we inspected the spatial
distribution of sources that have luminosities in the excess bins; no
concentration towards Trumpler 14 was found.

Coincidentally, \citet{C02} noted an apparent gap in the optical
color-magnitude diagram of Trumpler 15, similar to gaps observed for
some other open star clusters (e.g., Rachford \& Canterna 2000; Kumar
et al.\ 2008).  If the gap indeed indicates missing stars with
inferred main-sequence spectral type B1--B5 ($7-9\msun$; Girardi et
al. 2000), this anomaly could contribute to the observed deficit in
the XLF.  However, the origin of these optical color-magnitude gaps is
not yet clear: alternative explanations include the abrupt onset of
convection in the stellar envelopes \citep{BC74} and peculiarities in
the Balmer lines (Mermilliod 1976).

\subsection{Near-IR Color-Color and Color-Magnitude Diagrams}\label{sec:NIR}

We study the near-IR characteristics of Trumpler 15 stars using the
sub-sample of 738 X-ray selected sources that have valid $JHKs$ HAWK-I
photometry in all three bands.  Figure~\ref{fig:ccd} shows the near-IR
$J-H$ vs.\ $H-K$ color-color diagram.  The colors of most {\em
  Chandra} sources (concentrated at J-H=0.8, H-K=0.3) are consistent
with diskless young stars subjected to $A_V\sim 1-2$~mag (assuming
late type stars).

Following the same K-excess definition as \citet{P10}($> 0.05$ mag to
the right and below the reddened main sequence locus), our sample
contains 28 K-band excess stars, yielding a near-IR excess fraction of
$3.8\pm 0.7\%$ in Trumpler 15. The error on the $K$-excess fraction is
calculated using 1-$\sigma$ Poisson errors on the number of
stars. This is slightly higher than the fraction ($2.1\pm 0.7\%$)
presented in \citep{P10}, which used a different stellar sample (436
sources in the CCCP-Cl\#8 cluster that have near-IR photometry).  Both
estimates indicate that the Trumpler 15 excess fraction is lower than
in Trumpler 16 and Trumpler 14 which are $6.9\pm 1.3 \%$ and $9.7\pm
1.6\%$ \citep{P10}, respectively.  \citet{Wolk11} found a range of
K-excess disk fractions among the Trumpler 16 subclusters (e.g., $4\pm
3\%$ in subcluster 9 and $18\pm 4\%$ in the South Extension).

Figure~\ref{fig:cmd} shows the near-IR $J$ vs.\ $J-H$ color-magnitude
diagram (CMD) for the same stars shown in Figure~\ref{fig:ccd}. Known
OB stars are located at the top, reddened from the ZAMS with $A_V\sim
1$~mag.  The X-ray observations do not identify any previously unknown
$J < 10$ members of the cluster; the same result was found for other
unobscured clusters that are well studied in the optical \citep[e.g.,
  NGC 2244;][]{Wang08}.  The deficit of stars in the range $10<J<12$
seen in the diagram is attributed to the known lack of X-ray emission
from intermediate mass stars \citep[e.g.,][]{Stelzer06}.

The brightest stars ($J<14$) in the upper part of Figure~\ref{fig:cmd}
are bracketed by the 2 Myr (black line) and the 10 Myr (cyan line)
isochrones.  An age of 5-10 Myr for the low-mass stars is inferred
assuming the visual extinction $A_V\sim 1-2$ mag.  While \citet{C02}
suggested a cluster age between 2 Myr and 6 Myr for Trumpler 15,
\citet{T03} found that it is not possible to fit the optical CMD of
Trumpler 15 using a single age; a range of ages between 4 Myr and 30
Myr is required.  All these studies indicate that Trumpler 15 is more
evolved than Trumpler 14 (1--2 Myr; Vazquez 1996) and Trumpler 16
\citep[3--4 Myr;][]{P10}, a finding consistent with the lower near-IR
excess fraction discussed above.

Note that the spectral type of HD 93249 (O9III), the current most
massive member in Trumpler 15, also indicates that its stellar
population is likely more evolved than the other two clusters.
Although the lifetimes of massive stars are difficult to calculate
(and depend on parameters such as rotation, metallicity, and binarity
that cannot be known for exploded stars), recent stellar evolution
models indicate that stars more massive than 40 M$_\odot$ will produce
SNe within 5 Myr (Hirschi et al. 2004).  For 20 M$_\odot$ stars, it
takes less than 10 Myr.  Given Trumpler 15's rich population
(Section~\ref{sec:xlf}), there probably were stars formed with
sufficient mass ($\sim 20-30M_{\odot}$) to become Wolf-Rayet (WR)
stars, after passing through a blue supergiant (BSG) and either LBV or
maybe RSG phase (see Crowther 2007 for a review).  The lack of WR
stars in Trumpler 15 at the current epoch \citep{S06} and the inferred age from the near-IR
CMD suggest that the more massive stars have evolved through the WR
stage to SN.  Although earlier studies did not find much evidence for
a recent SN in the Carina complex \citep{SB08, Ezoe09, Townsley11b},
the CCCP provides several lines of evidence in favor of SNe from an
older generation of stars: a young X-ray-emitting neutron star
(Hamaguchi et al. 2009), strong metal-enriched X-ray plasma (Townsley
et al. 2011b), and a widely dispersed population of both PMS and OB
stars (Feigelson et al. 2011).  It is also interesting to note that
there are one BSG (Trumpler 15 18) and two RSGs (RT Car and BO Car) in
the Trumpler 15 field, which could be evolved massive members.
Currently their membership to Trumpler 15 is inconclusive
\citep{M88,T03,S06}.

\section{X-rays from Individual Stars}\label{sec:individual}

Seven X-ray sources in Trumpler 15, listed in Table 3, are identified
as counterparts \citep{Broos11b} to stars with OB spectral types
(Morrell et al.\ 1988; Carraro 2002; Skiff 2009).  More comprehensive
investigations of massive stars are available in CCCP studies by
\citet{Gagne11} and by \citet{Naze11}.

The visually bright star ($V=8.4$) HD 93249 (=Cl Trumpler 15
1=CD-58$^{\circ}$3536) in the center of Trumpler 15 has the earliest
spectral type, O9 III (Morrell 1988; Walborn 1973; but see also
Feinstein et al. 1980 who assigned a spectral type O8 II,m).  The
spectrum (184 X-ray counts) extracted from this source
(Figure~\ref{fig:Ospec}) is well fit ($\chi^2_{\nu}=0.9$) using XSPEC
(version 12.6, Arnaud 1996) with a $kT=0.36^{+0.20}_{-0.10}$ keV
thermal plasma model (APEC; Smith et al. 2001) absorbed by $N_H=4.8\pm
2.5\times 10^{21}$ cm$^{-2}$ ($tbabs$; Wilms et al. 2000). The
elemental abundances were fixed at $Z = 0.3Z_{\odot}$ for the spectral
fit. This choice of subsolar abundances follows the COUP study of
Orion OB stars (Stelzer et al.\ 2005). The derived X-ray absorption
gives an $A_V=3\pm 1$ mag (assuming $N_H/A_V=1.6\times 10^{21}$
cm$^{-2}$; Vuong et al. 2003), consistent with the average optical
extinction towards Trumpler 15 \citep[$A_V\approx 2$ mag;][]{C02}. The
absorption corrected total band (0.5–-8 keV) X-ray luminosity,
$L_{t,c}$, is $7.4\times 10^{31}$ erg s$^{-1}$.  Adopting $\log
(L_{bol}/L_{\odot})=5.28$ for this spectral type (see Table 4 in Smith
2006), we find $L_x/L_{bol}=1.0\times 10^{-7}$, which nicely follows
the standard empirical relation \citep[e.g.,][]{SC82,Sana06}. This
star has a low-temperature X-ray plasma consistent with the standard
model of shocks within an unstable radiatively driven wind
\citep{Lucy80}.

We detect a weak X-ray source (17 counts) 7\arcsec\/ north of HD
93249, consistent with the position reported for its companion
CD-58$^{\circ}$3536B (=Cl Trumpler 15 2; O9-9.5 III,nn; Morrell 1988);
no analysis of its spectrum was attempted.

Among the census of O stars in Trumpler 15 \citep{C02,S06}, we do not
detect X-ray emission from two early-type stars: HD 93342 (O9 III),
listed as an uncertain member by Feinstein et al. (1980), and a BSG
Trumpler 15 18 (O9 I; Morrell et al. 1988).  The latter suffers an
additional $A_V=3$ mag of dust extinction compared to the other
cluster members \citep{T03}.  If we assume a soft thermal spectrum
with temperature of $kT\sim 0.4$ keV as in HD~93249, an absorption
column $N_H\sim 10^{22}$ cm$^{-2}$, and an X-ray luminosity of $L_x
\sim 1.5 \times 10^{32}$ erg s$^{-1}$ based on the $\log L_x/L_{bol}
\approx -7$ relation, then we predict about $\sim 80$ counts in the
CCCP data from Trumpler 15 18.  Therefore the
non-detection\footnote{The full ACIS Extract \citep{Broos10} procedure
  used in the CCCP estimates the probability $ProbNoSrc$ that the
  source counts were due to fluctuations in the observed local
  background in three bands (soft [0.5--2 keV], hard [2--8 keV], and
  total [0.5--8 keV]), taking into account the variation of {\em
    Chandra} point spread function with off-axis angle and the
  presence of nearby X-ray point sources. A detection requires both
  the total source counts $SrcCounts\_t\ge 3$ and the smallest
  $ProbNoSrc$ in the three band $< 0.01$. See Broos et al.(2011a) for
  details of CCCP source detection.} of this star with an upper limit
of a few counts supports a note made in Morrell et al. (1988) that it
is probably an evolved background star.  A similar calculation applies
for HD 93342.  Thus, the X-ray evidence suggests that neither HD 93342
nor Trumpler 15 18 are massive Carina members, unless they are
obscured by high absorption columns.

Another X-ray star ,CCCP \#7584 (CXOGNC J104446.54-592154.0), is worth
noting, although it is detected with only 7 net counts. The near-IR
counterpart is Cl Trumpler 15 10 (B2V), showing significant K-band
excess ($J-H=-0.14$ and $H-K=0.4$ in Figure~\ref{fig:ccd}).  It is
also identified as a young stellar object showing mid-IR excess
emission (source \#670; Povich et al. 2011).  This high mass star may
still possess an optically thick inner accretion disk and belong to
the class of Herbig Be stars (Herbig 1960).

\section{Summary}\label{sec:sum}

The $Chandra$ ACIS observations of the Trumpler 15 cluster are
presented in the framework of the CCCP survey.  The results are
summarized as follows:

1. We provide an X-ray source list of the Trumpler 15 region,
containing 829 probable Carina members with an X-ray luminosity
completeness limit $\log L_{t,c}\sim 30.4$ erg s$^{-1}$.  Positional
coincidence matching yields a total of 748 HAWK-I near-IR
counterparts.  This represents an increase in membership by a factor
$>30$ over previous optical studies that were mostly limited to OB
stars.

2. The X-ray detected population provides the first deep probe of the
rich low mass population in this cluster.  The projected density of
Trumpler 15 stars indicates a compact and spherical structure within
the inner 1\arcmin\/ (0.7 pc) region, and an elongation along the N-S
direction at larger distances ($>4\arcmin$) from the core.  A
continuous ``envelope'' of dispersed stars connects Trumpler 15 with
Trumpler 14 to the SW, affirming the view that the two clusters lie at
the same distance.  The radial density profile of the X-ray cluster is
sharply peaked in the inner $r<1\arcmin$ region, and the massive stars
are spatially concentrated towards the cluster center.  In contrast,
the PMS stars with dusty disks, identified by near-IR
excess, are distributed throughout the cluster.

3. We interpret the XLF ($L_{t,c}$) for our Trumpler 15 sources as a
standard XLF shape (defined by the COUP ONC) that is incompletely
detected for $\log L_{t,c} \lesssim 30.4$ erg s$^{-1}$ and that
exhibits a deficit of sources for $\log L_{t,c} \gtrsim 31.0$ erg
s$^{-1}$.  This deficit of luminous X-ray sources in the current epoch
is best explained by the destruction of massive stars in the original
population as they evolved into supernovae.  Comparison of the two
XLFs within a restricted luminosity range where they have similar
slopes indicates that Trumpler 15 has $\sim$5900 stars ($2.1 \pm 0.3$
times the ONC population).

4. The near-IR color-magnitude diagram of X-ray selected sources
indicates an age of 5--10 Myr and a K-excess disk frequency of $3.8\pm
0.7\%$ for the PMS population.  This age estimate, the absence of WR
stars and main sequence members earlier than O9, and the unusual
diffuse X-ray emission in the Carina complex all suggest that the most
massive original members of the cluster ($>20-40$ M$_\odot$) have
evolved into supernovae.

5. We detected seven massive stars with spectral types ranging from B2
to O9.  HD 93249 (O9 III) has the soft spectrum expected for X-ray
emission via wind shocks in a massive star. Its X-ray luminosity
($L_{t,c}=7.4\times 10^{31}$ erg s$^{-1}$) is also consistent with the
long-standing $\log(L_x/L_{bol})\sim -7$ relation for O stars.  The
X-ray source Cl Trumpler 15 10 (B2) is proposed as a candidate Herbig
Be star due to its significant K-band excess.

We thank the anonymous referee for providing constructive comments,
M. Povich for helpful discussion, and G. Carraro for information on
the optical catalog published in Carraro (2002).  This work is
supported by {\em Chandra} X-ray Observatory grant GO8-9131X (PI:
L.\ Townsley) and by the ACIS Instrument Team contract SV4-74018 (PI:
G.\ Garmire), issued by the {\em Chandra} X-ray Center (CXC), which is
operated by the Smithsonian Astrophysical Observatory for and on
behalf of NASA under contract NAS8-03060.  J.W. is supported by NASA
grant NNX08-AW89G and {\em Chandra} grant GO8-9101X (PI: Fabbiano).
S.J.W. is supported by NASA contract NAS8-03060 (CXC). A.F.J.M thanks
NSERC (Canada) and FQRNT (Quebec) for financial assistance. R.R.K. is
supported by a Leverhulme research project grant (F/00 144/BJ). The
near-infrared observations were collected with the HAWK-I instrument
on the VLT at Paranal Observatory, Chile, under ESO program
60.A-9284(K).  This publication makes use of data products from the
Two Micron All Sky Survey, which is a joint project of the University
of Massachusetts and the Infrared Processing and Analysis
Center/California Institute of Technology, funded by NASA and the
National Science Foundation.  This research has made use of the SIMBAD
database and the VizieR catalogue access tool, operated at CDS,
Strasbourg, France.

{\it Facilities:} \facility{CXO (ACIS)}

%-----
\clearpage

\begin{figure}[h]
\centering 
%\epsscale{0.5}
\plotone{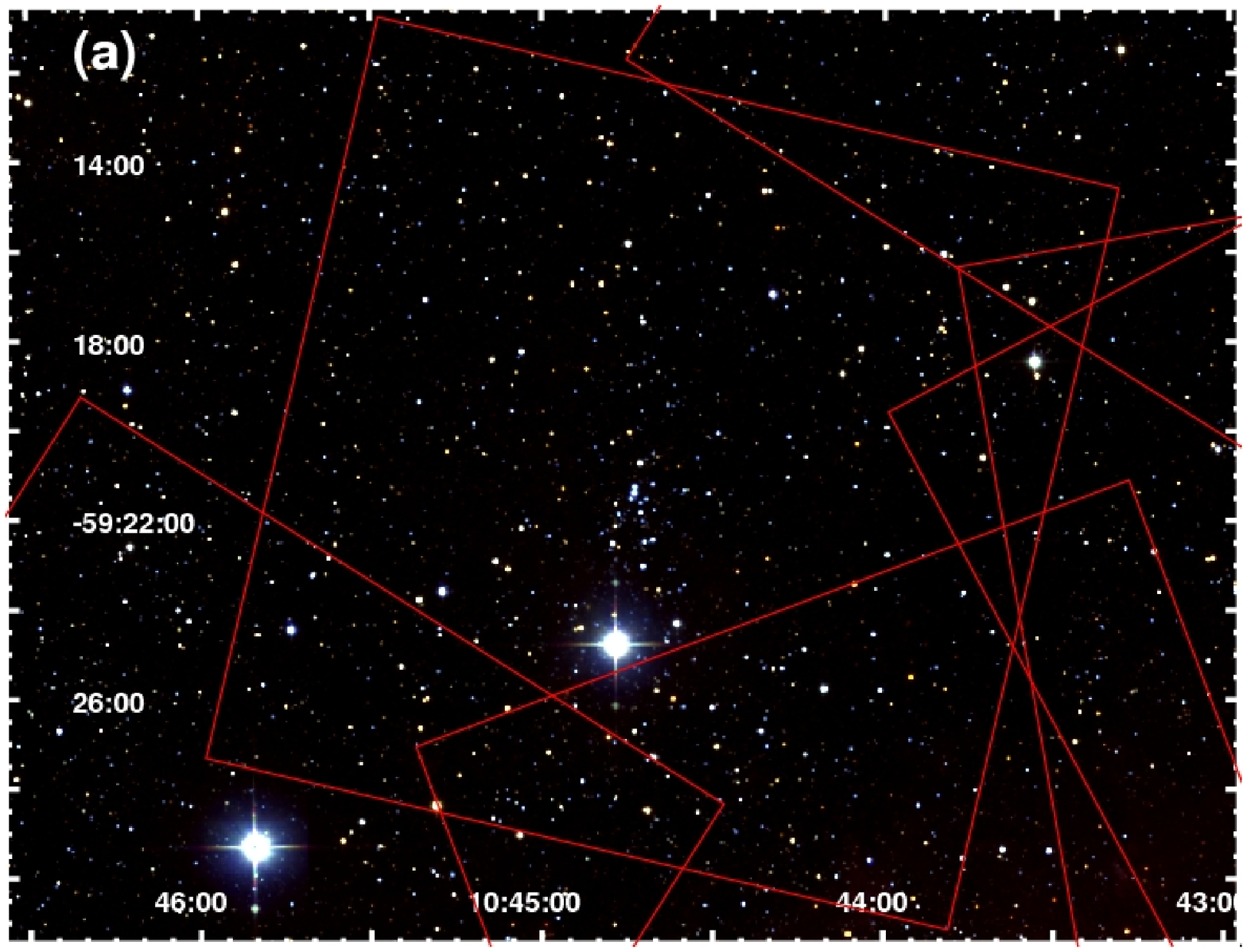}
%\epsscale{1.0}
\caption{(a) 2MASS $JHK_s$ composite image of the Trumpler 15 region overlaid with boxes depicting the $17\arcmin \times 17\arcmin$ ACIS-I fields of view for the CCCP observations.\label{fig:fig2}}
\end{figure}

\addtocounter{figure}{-1} 
\begin{figure}[h]
\centering 
\plotone{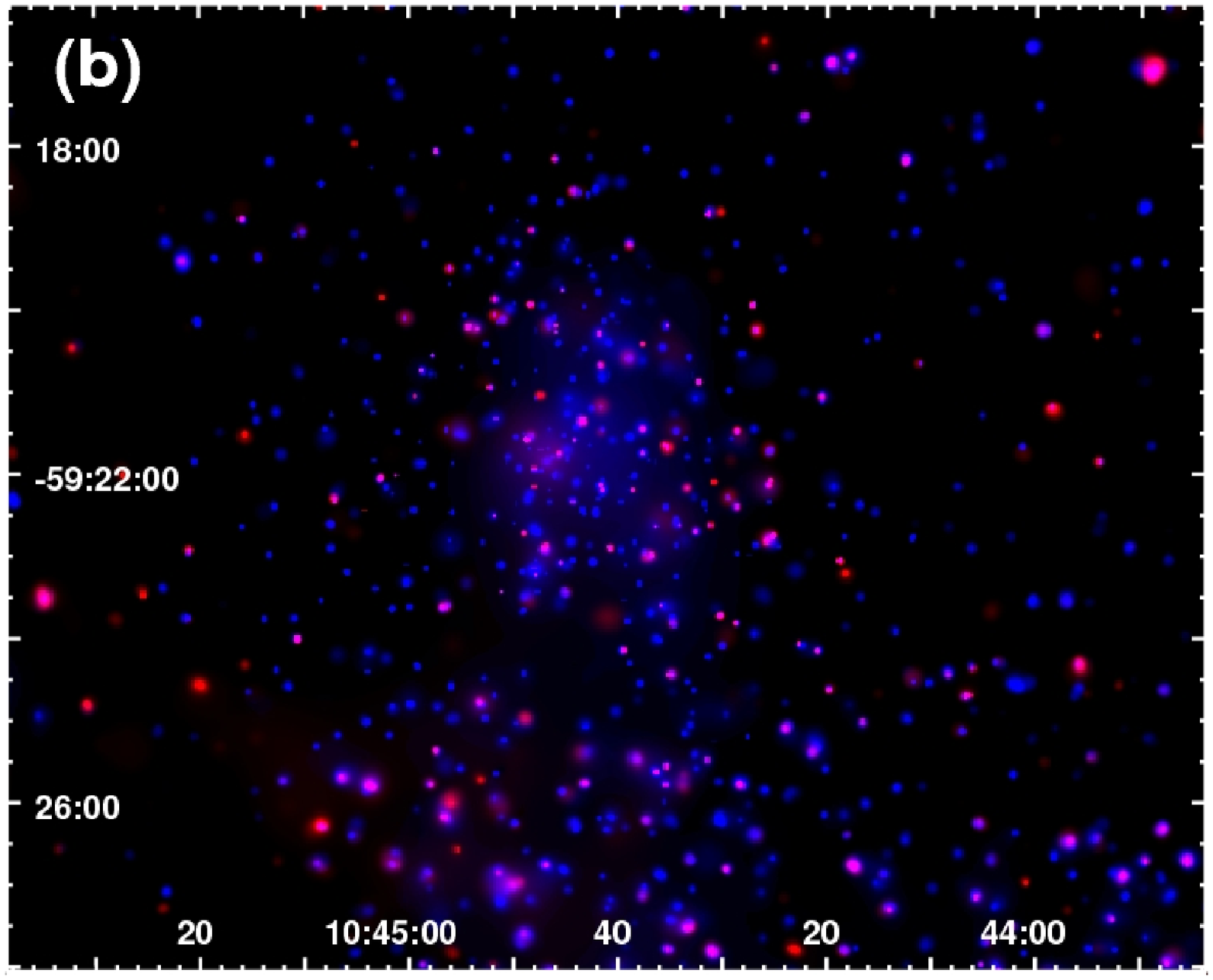}
\caption{(b) Adaptively smoothed {\em Chandra} ACIS images of Trumpler
  15 ($\sim 12\arcmin\times 12\arcmin$). Red represents the soft-band
  (0.5--2 keV) X-ray emission and blue the hard X-ray emission (2--7
  keV).  The sky coordinates are for the epoch J2000 in both panels.}
\end{figure}

\begin{figure}[h]
\centering \plotone{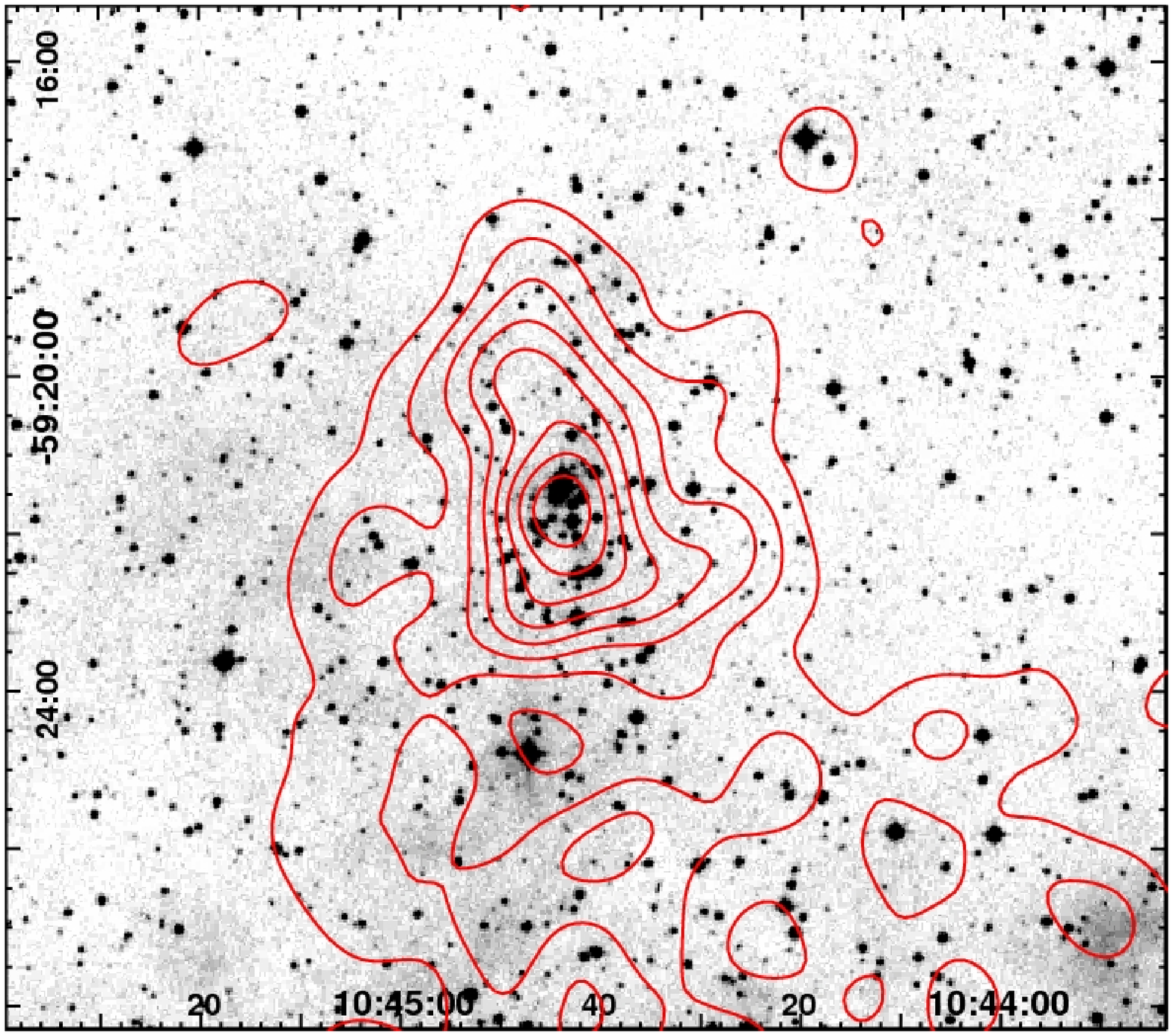} \caption{DSS $R$-band image of the
  Trumpler 15 region overlaid with X-ray source density contours
  \citep{F10}. The sky coordinates are for the epoch
  J2000. \label{fig:fig1}}
\end{figure}

\begin{figure}[h]
\centering \plotone{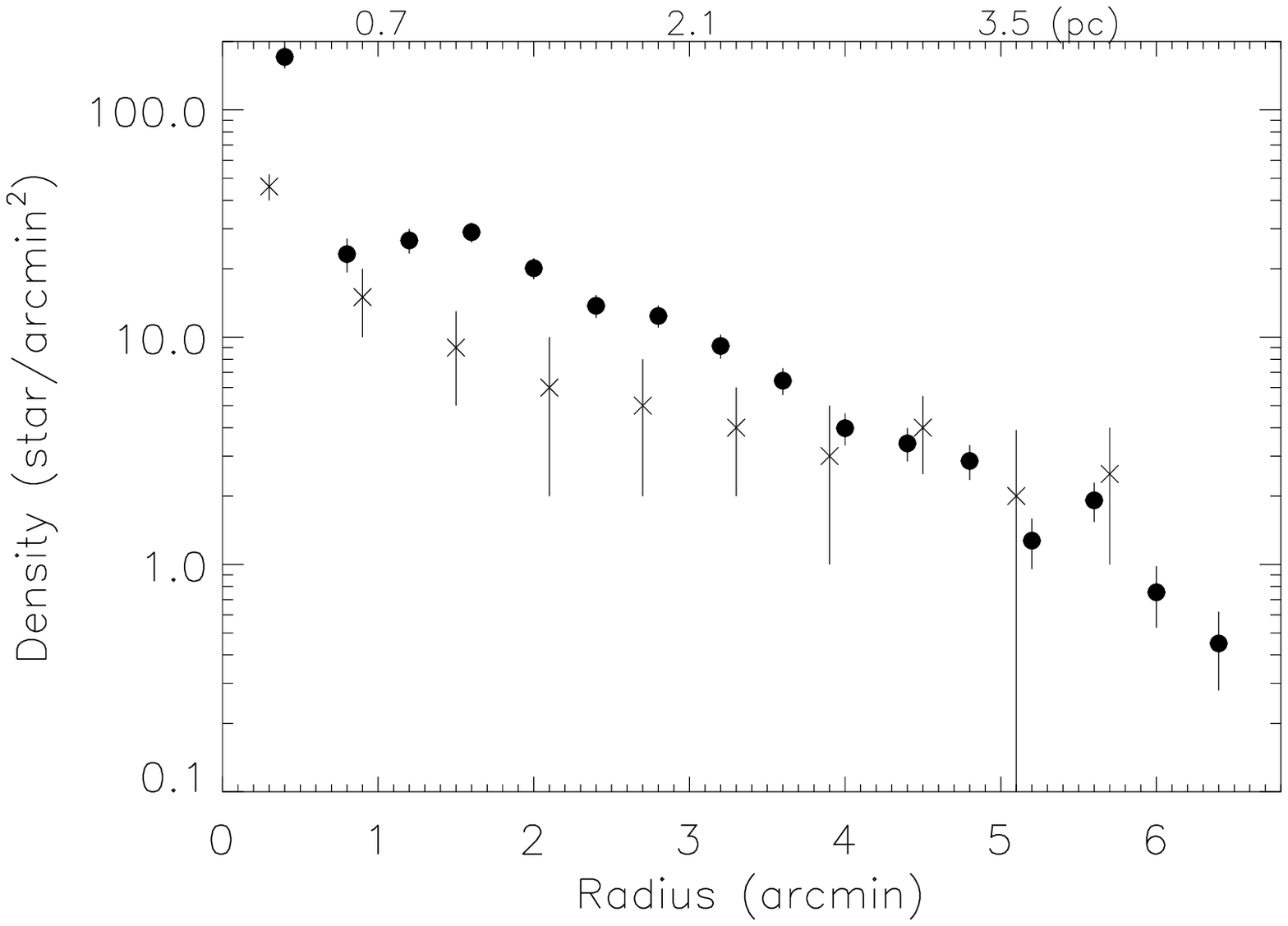} \caption{Radial density profile of X-ray sources (filled circles) and optical sources (crosses, Carraro 2002).\label{fig:fig3}}
\end{figure}

\begin{figure}[h]
\centering
\plotone{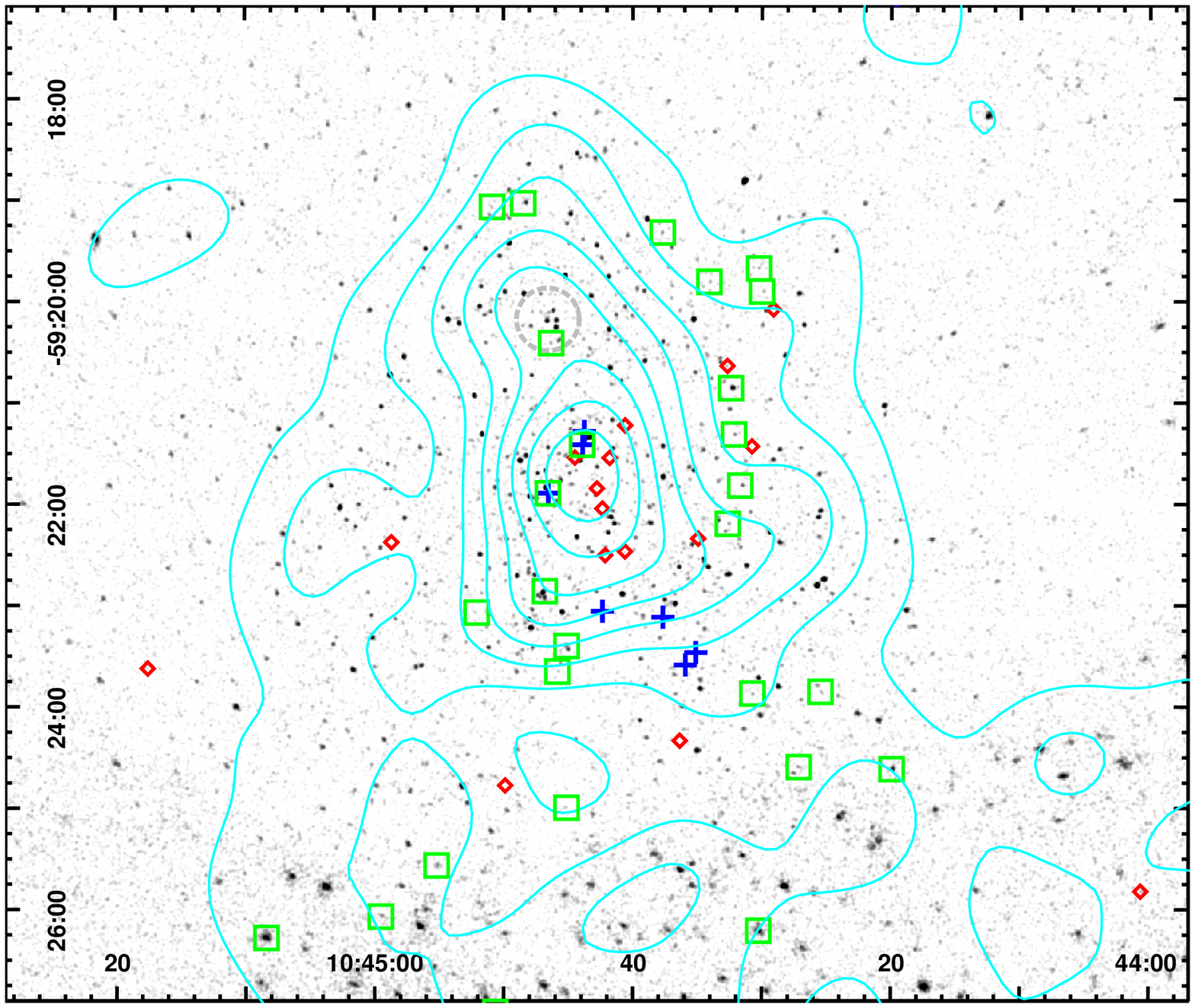}
\caption{(a) Spatial distribution of three groups of stars relative to
  lower mass cluster members detected by the CCCP: X-ray detected
  (blue crosses) and undetected high mass stars (red diamonds) in
  Skiff (2009), and X-ray stars with K-band excess (green boxes).
  Density of cluster members is shown as contours (cyan) on a 0.5--7
  keV X-ray image.  The location of a small clump of stars noted in
  the text is marked with a grey circle.\label{fig:himass}}
\end{figure}

\addtocounter{figure}{-1} 
\begin{figure}[h]
\centering 
\plotone{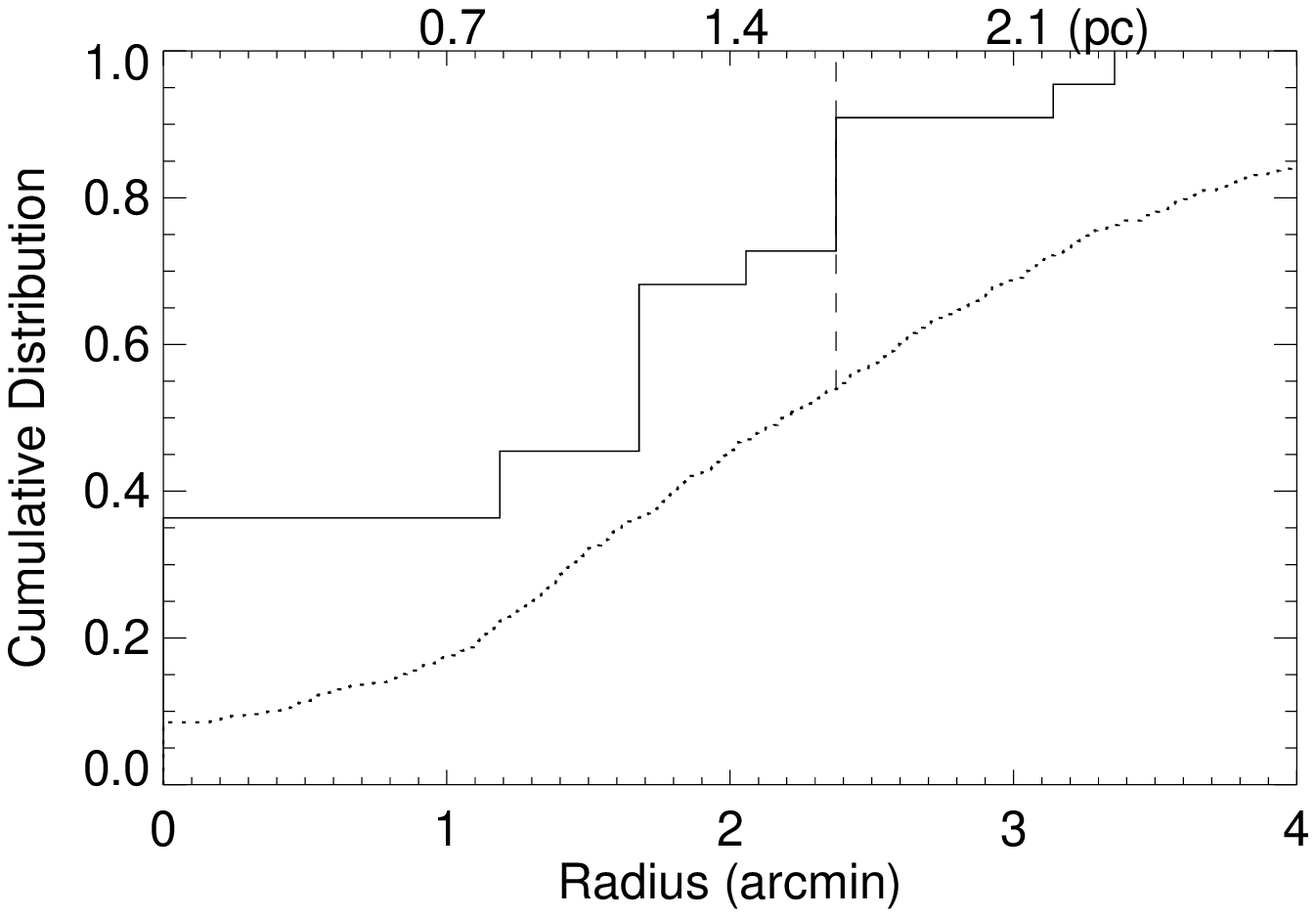}
\caption{(b) One-dimensional cumulative radial distributions for all
  high mass stars (solid) and for the lower mass CCCP cluster members
  (dotted).  Using a 2-sample Kolmogorov-Smirnov statistic computed at
  the dashed line, we find a probability of 0.4\% that the two
  populations have the same radial profile.}
\end{figure}

\begin{figure}[h]
\centering \plotone{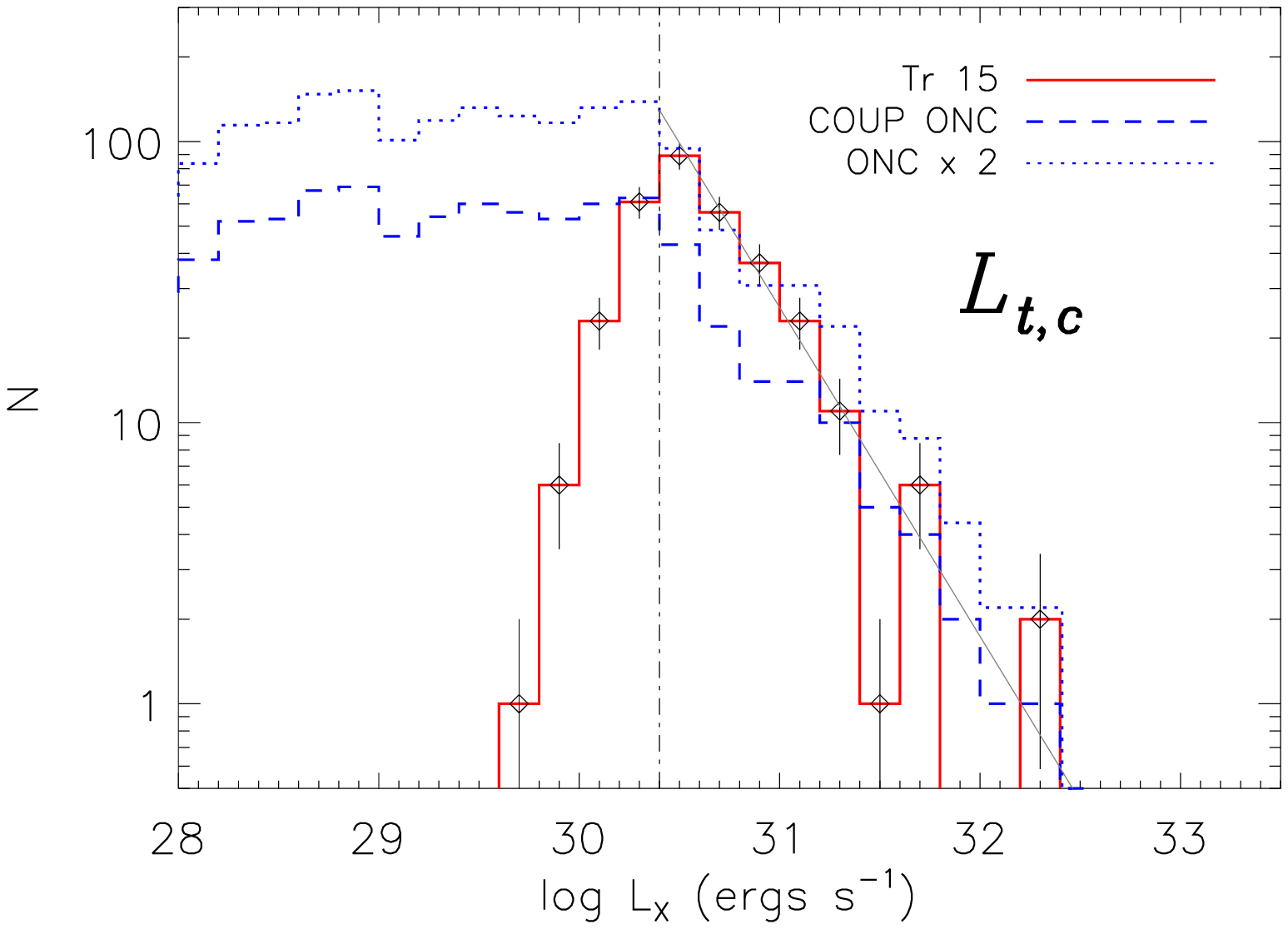} \caption{Distributions of total-band
  absorption corrected X-ray luminosity for 316 low-mass CCCP stars in
  Trumpler 15 (red solid line, 7 known OB stars omitted), and for 839
  low-mass COUP stars in the ONC (blue dashed line).  The blue dotted
  line shows the COUP XLF scaled by a factor of 2 to match Trumpler 15
  for $30.4\leq \log L_{t,c}\leq 31.0$. The grey solid line represents
  a fit with a power-law slope $\Gamma=-1.27$, and the vertical
  dash-dotted line is the estimated completeness limit for
  $L_{t,c}$.\label{fig:XLF}}
\end{figure}

\begin{figure}[h]
\centering \epsscale{0.8} 
\plotone{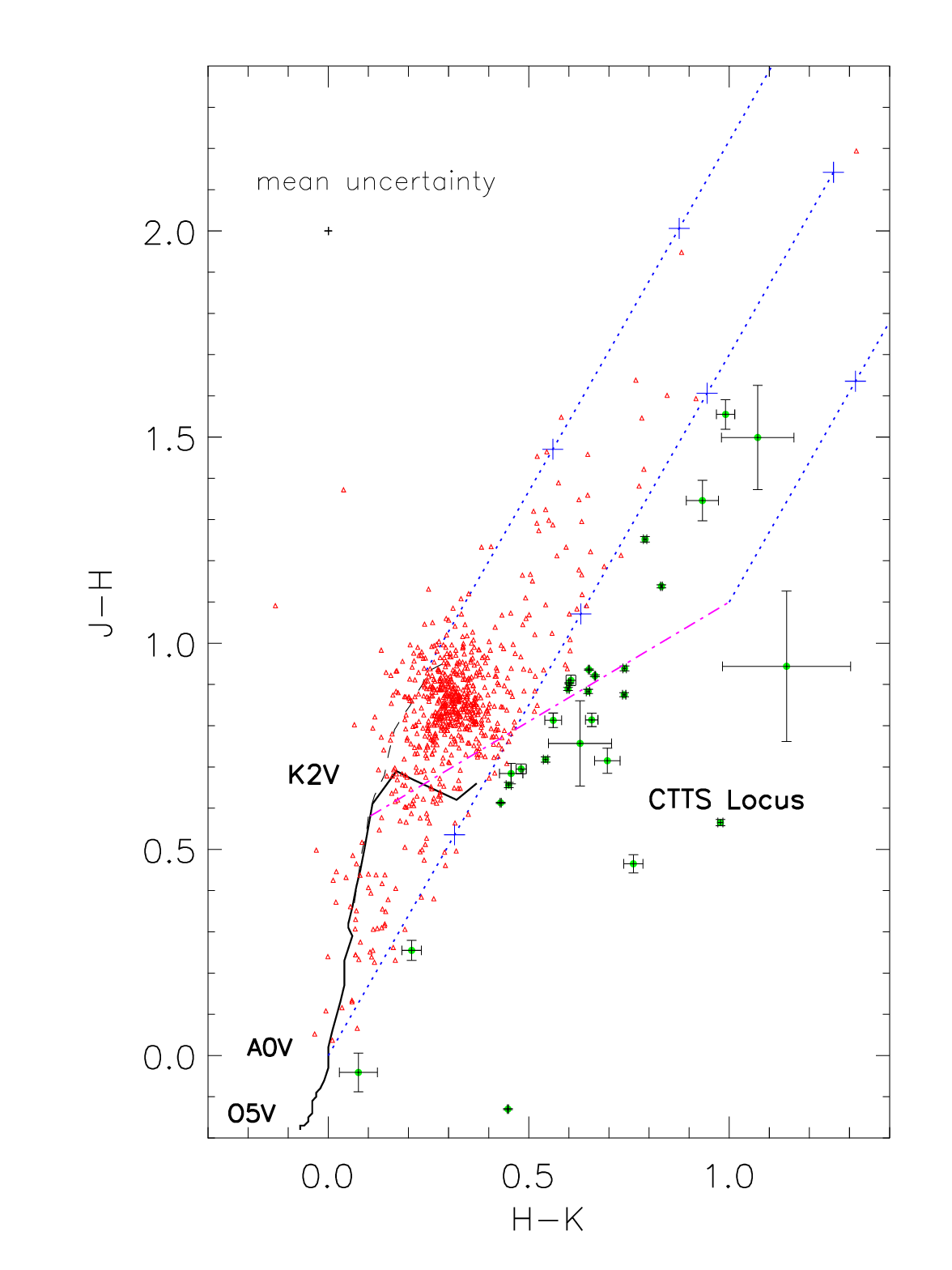}
\epsscale{1.} \caption{Near-IR color-color diagram ($J-H$ vs. $H-K$)
  for the X-ray selected Trumpler 15 probable members.  The green
  circles and red triangles represent sources with and without
  significant $K$-band excess.  The black solid and long-dash lines
  denote the loci of Zero Age Main Sequence (ZAMS) stars and giants,
  respectively, from Bessell \& Brett (1988). The purple dash dotted
  line is the locus for classical T Tauri stars (CTTS) from Meyer et
  al. (1997). The blue dashed lines represent the standard reddening
  vector with crosses marking $A_V$ intervals of 5
  magnitudes.\label{fig:ccd}}
\end{figure}

\begin{figure}[h]
\centering 
\epsscale{0.8}
\plotone{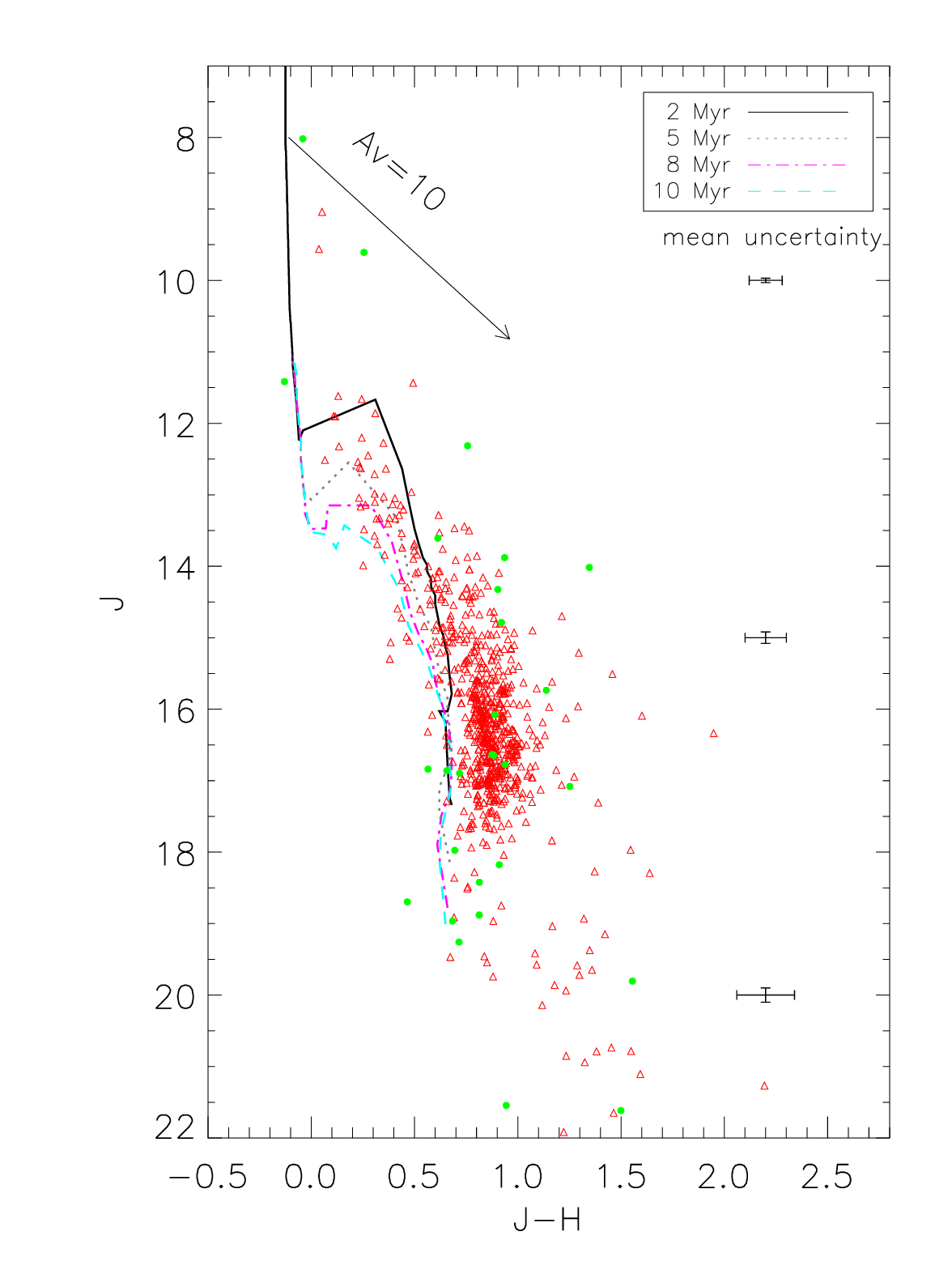} 
\epsscale{1.0}
\caption{Near-IR color-magnitude diagram ($J$ vs. $J-H$) using the
  same Trumpler 15 sample and symbols as in Figure~\ref{fig:ccd}. The
  solid (black) line shows the 2 Myr isochrone; the dotted (grey), the
  dot-dashed (magenta), and the dashed (cyan) lines show the low-mass
  ($M\leq 7 \msun$) isochrones \citep{Siess00} for ages of 5, 8, and
  10 Myr, respectively.  The arrow indicates the redenning vector for
  $A_V$ = 10 mag. Typical uncertainties of the photometry are
  indicated by the three error bars to the right. \label{fig:cmd}}
\end{figure}

\begin{figure}[h]
\centering
\includegraphics[scale=.6,angle=-90]{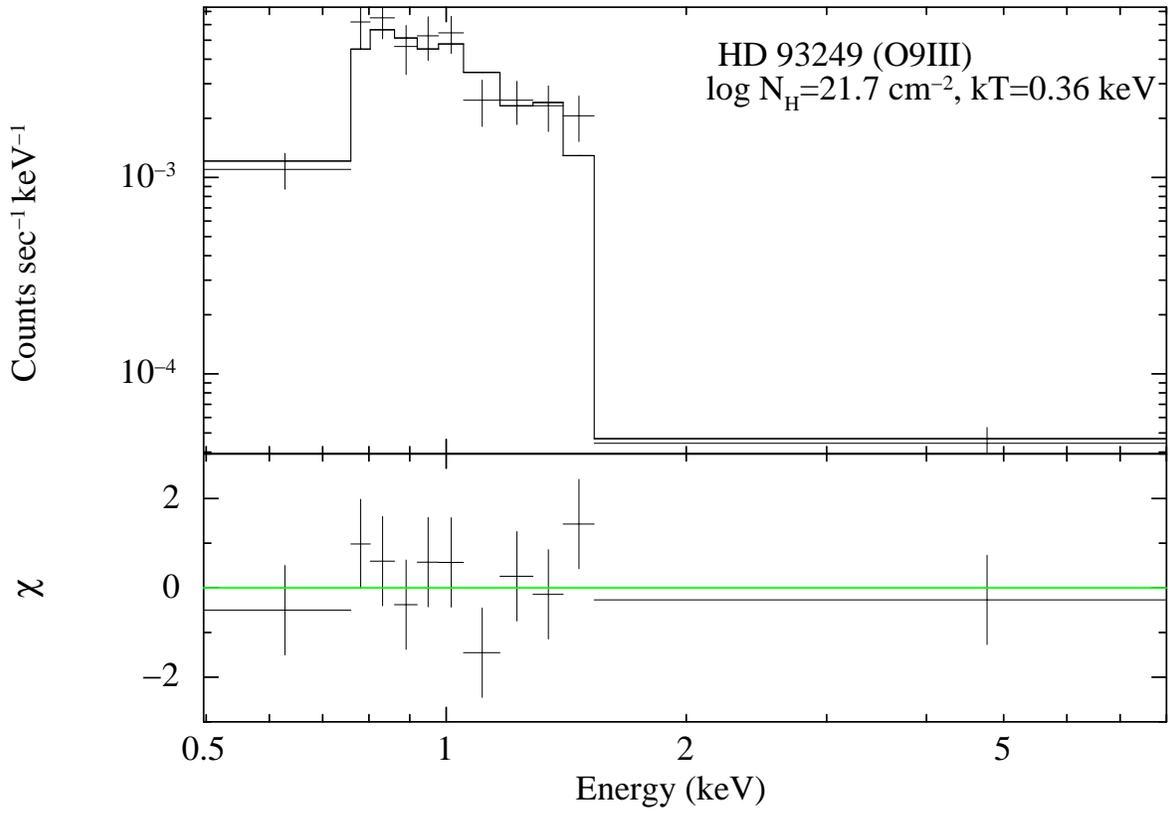} \caption{X-ray spectrum
  of the O9III star HD 93249 and the best fit with a simple absorbed
  thermal plasma model.\label{fig:Ospec}}
\end{figure}

\clearpage
%---------------
%\pagestyle{empty}
\begin{deluxetable}{lcccc}
\centering
%\tabletypesize{\small}
\tablewidth{0pt} \tablecolumns{5}

\tablecaption{Properties of Trumpler 15 in the Literature
\label{tbl:basic}} \tablehead{\colhead{Reference} &
\colhead{Distance} &
\colhead{Age} & \colhead{Members} & \colhead{E(B-V)}\\
\colhead{}  & \colhead{(kpc)} & \colhead{(Myr)} & \colhead{(\#))} & \colhead{(mag)}\\
}
\startdata
1& 1.7  & ...  &    10  &   0.53 \\
2& 1.6  &  2   &    25  &   0.4  \\
3& 3.7  & ...  &   ...  &   ...  \\   
4& $2.6\pm 0.2$ &  $6\pm 3$  & 36 & $0.48\pm0.07$ \\
5& 2.5  & ...   &  ...   &  0.189 \\
6& $2.4\pm0.2$ & ...  &  35    &   $0.49\pm 0.09$\\
7& $2.4\pm0.3$ & $<$6 &  90    &   $0.52\pm 0.07$\\
8& 2.9 & 3-40 &  ...  & $>$0.42\\
\enddata

\tablecomments{References--[1] \citet{G68}; [2] \citet{T71}; [3] \citet{W73,W95}; [4]
\citet{F80}; [5] \citet{T80}; [6] \citet{T88}; [7] \citet{C02};
[8] \citet{T03}}

\end{deluxetable}

\clearpage
%---------------
%\pagestyle{empty}
\begin{deluxetable}{lccccc}
\centering
%\tabletypesize{\small}
\tablewidth{0pt} \tablecolumns{6}

\tablecaption{Properties of Chandra ACIS Point Sources in the Trumpler 15 Region \label{tbl:list}} 
\tablehead{\colhead{Seq. \#} & \colhead{Designation} &
\colhead{R.A.} &
\colhead{Dec.} & \colhead{NetCounts\_t} & \colhead{Class}\\
\colhead{}  & \colhead{(CXOGNC J)} & \colhead{(J2000)} & \colhead{(J2000))} & \colhead{} & \colhead{}\\
}
\startdata
   5823&104416.56-592501.3&161.069031&-59.417038& 38.3&H2\\
   5930&104418.41-592608.2&161.076736&-59.435612& 12.4&H2\\
   5968&104419.32-592209.6&161.080521&-59.369358&  6.6&H2\\
   5989&104419.66-592222.3&161.081924&-59.372875& 17.6&H2\\
   5992&104419.72-592359.4&161.082184&-59.399860&  3.3&H2\\
   6002&104419.88-592437.0&161.082840&-59.410294& 30.6&H2\\
   6023&104420.20-592534.7&161.084198&-59.426323& 16.6&H2\\
   6027&104420.25-592608.8&161.084396&-59.435799& 22.1&H2\\
   6043&104420.47-592402.2&161.085327&-59.400627&  3.4&H2\\
   6053&104420.72-592433.8&161.086365&-59.409397&  2.8&H2\\
\enddata

\tablecomments{Table \ref{tbl:list} with complete notes is
  published in its entirety in the electronic edition of the {\it
    Astrophysical Journal}.  A portion is shown here for guidance
  regarding its form and content.}

\tablecomments{{\bf Column 1:} CCCP X-ray catalog sequence number
  \citep{Broos11a}.  {\bf Column 2:} IAU designation.  {\bf Columns
    3,4:} Right ascension and declination (both in degrees) for epoch
  J2000.0.  {\bf Column 5:} Net X-ray events detected in the source
  extraction aperture in the total band (0.5--8 keV), taken from Table
  1 in \citet{Broos11a} with the identical column name. {\bf Column
    6:} A set of mutually exclusive classification hypotheses defined
  for each source in \citet{Broos11b}---H0: source is unclassified;
  H1: source is a foreground main-sequence star; H2: source is a young
  star, assumed to be in the Carina complex; H3: source is a Galactic
  background main-sequence star; H4: source is extragalactic.}

\tablecomments{Notes/other names from SIMBAD queries using a 2\arcsec\/ matching radius:}

\tablenotetext{6860}{---Cl Trumpler 15 20=GSC 08626-00886=GEN\# +4.35150020}
\tablenotetext{6907}{---Cl Trumpler 15 19=Cl* Trumpler 15 C 28=GEN\# +4.35150019}
\tablenotetext{7023}{---Cl Trumpler 15 21=Cl* Trumpler 15 C 44=GEN\# +4.35150021}
\tablenotetext{7195}{---Cl Trumpler 15 8=GSC 08626-01516=GEN\# +4.35150008}
\tablenotetext{7318}{---Cl Trumpler 15 15=CPD-58 2656=Cl* Trumpler 15 C 3= GSC 08626-01303=GEN\# +4.35150015=TYC 8626-1303-1}
\tablenotetext{7332}{---Cl Trumpler 15 6=NSV 18508}
\tablenotetext{7397}{---Cl Trumpler 15 2=CCDM J10447-5921B=Cl* Trumpler 15 C 2=GSC 08626-02845=GEN\# +4.35150002 }
\tablenotetext{7403}{---Cl Trumpler 15 1=HD 93249=ALS 1857=1E 1042.7-5905=SAO 238421=CCDM J10447-5921A=GEN\# +4.35150001J}
\tablenotetext{7584}{---Cl Trumpler 15 10=Cl* Trumpler 15 C 17 =CPD-58 2662=   	  GSC 08626-01107  =GEN\# +4.35150010 }
\tablenotetext{7679}{---Cl Trumpler 15 12=GSC 08626-01694 =   	  GEN\# +4.35150012  }
\tablenotetext{7696}{---Cl Trumpler 15 38 =   	  GEN\# +4.35150038=2MASS J10444861-5922153  }
\tablenotetext{7756}{---Cl Trumpler 15 29 =  GSC 08626-01760 	 = GEN\# +4.35150029  }
\tablenotetext{8235}{---Cl Trumpler 15 31 = GEN\# +4.35150031 }
\tablenotetext{8503}{---Cl Trumpler 15 35 = GEN\# +4.35150035=2MASS J10450167-5923382}
\tablenotetext{8554}{---Cl Trumpler 15 34 = GEN\# +4.35150034=2MASS J10450269-5922019}
\end{deluxetable}

\clearpage
%---------------
%\pagestyle{empty}
\begin{deluxetable}{cccccccc}
\centering
\tabletypesize{\footnotesize}
\tablewidth{0pt} \tablecolumns{8}

\tablecaption{CCCP Detected OB Stars in the Trumpler 15 Region \label{tbl:detect}} 
\tablehead{\colhead{Name} & \colhead{CCCP} &
\colhead{R.A.} &
\colhead{Dec.} & \colhead{Sp.Type} & \colhead{$V$} & \colhead{X-ray Cts} & \colhead{Note} \\
\colhead{Cl Trumpler 15}  &\colhead{ID}  & \colhead{(J2000)} & \colhead{(J2000))}  & \colhead{} & \colhead{(mag)}  & \colhead{(0.5--8 keV)} & \colhead{}\\
}
\startdata
 1   &  7403     &    161.18283    &    -59.35700  & O9III   &   8.40  &  184 & HD 93249; CD-58 3536 \\
 2    &  7397     &    161.18229    &    -59.35481  &  O9.5:III &  9.50 & 17 & CD-58 3536B\\
 10   &   7584    &     161.19387    &    -59.36497 &  B2V       &  11.60 &7 & CPD-58 2662\\
 15   &   7318    &     161.17642    &    -59.38439 &  B0.5IV-V &10.10 &  14& CPD-58 2656\\
 19   &   6907    &     161.14967    &    -59.39322  & O9:V:    & 12.70 &  12& ...\\
 20   &   6860     &    161.14629    &    -59.39117  & O9 V:       & 12.70 &  3& ... \\
 21    &  7023    &     161.15692     &   -59.38536 &  B0:III:  & 13.10 &  3& ...\\
\enddata

\tablecomments{{\bf Column 1:} Name of Trumpler 15 OB stars originally
  defined by \citet{G68}.  {\bf Column 2:} CCCP X-ray catalog sequence
  number \citep{Broos11a}.  {\bf Columns 3,4:} Right ascension and
  declination (both in degrees) for epoch J2000.0.  {\bf Column 5:}
  Spectral type as classified in \citet{Skiff09}. {\bf Column 6:} $V$
  magnitude \citep{C02}.  {\bf Column 7:} Net X-ray events detected in
  the source extraction aperture in the total band (0.5--8 keV), taken
  from Table 1 in \citet{Broos11a} with column name
  ``NetCounts\_t''. {\bf Column 8:} Other designations listed in
  SIMBAD.}
\end{deluxetable}


\begin{thebibliography}{99}

\bibitem[Arnaud(1996)]{1996ASPC..101...17A} Arnaud, K.~A.\ 1996, 
Astronomical Data Analysis Software and Systems V, 101, 17 

\bibitem[Ascenso et 
al.(2007)]{2007A&A...476..199A} Ascenso, J., Alves, J., Vicente, S., \& Lago, M.~T.~V.~T.\ 2007, \aap, 476, 199 

\bibitem[Bessell \& Brett(1988)]{Bessell88} Bessell, M.~S., \& Brett, J.~M.\ 1988, \pasp, 100, 1134

\bibitem[Bohm-Vitense \& Canterna(1974)]{BC74} Bohm-Vitense, E., \&
  Canterna, R.\ 1974, \apj, 194, 629

\bibitem[Broos et~al.(2007)]{Broos07}
{Broos}, P.~S., {Feigelson}, E.~D., {Townsley}, L.~K., {Getman}, K.~V., {Wang}, J., {Garmire}, G.~P., {Jiang}, Z., \& {Tsuboi}, Y. 2007, \apjs, 169, 353

\bibitem[Broos et al. (2010)]{Broos10} Broos, P. S., Townsley, L. K., Feigelson, E. D., Getman, K. V., Bauer, F. E., \& Garmire, G. P. 2010, \apj, 714, 1582

\bibitem[Broos et al.(2011a)]{Broos11a} Broos, P.~S., et al.\ 2011a, \apjs, submitted (CCCP Catalog Paper)

\bibitem[Broos et al.(2011b)]{Broos11b} Broos, P.~S., et al.\ 2011b, \apjs, submitted (CCCP Classifier Paper)

\bibitem[Carraro(2002)]{C02} Carraro, G.\ 2002, \mnras, 
331, 785 

\bibitem[Crowther(2007)]{2007ARA&A..45..177C} Crowther, P.~A.\ 2007, \araa, 45, 177 

\bibitem[Davidson 
\& Humphreys(1997)]{DH97} Davidson, K., \& Humphreys, R.~M.\ 1997, \araa, 35, 1 

\bibitem[Dias et 
al.(2002)]{2002A&A...389..871D} Dias, W.~S., Alessi, B.~S., Moitinho, A., \& L{\'e}pine, J.~R.~D.\ 2002, \aap, 389, 871 

\bibitem[Ezoe et al.(2009)]{Ezoe09} Ezoe, Y., Hamaguchi, K., 
Gruendl, R.~A., Chu, Y.-H., Petre, R., 
\& Corcoran, M.~F.\ 2009, \pasj, 61, 123 

\bibitem[Feinstein et al.(1980)]{F80} Feinstein, A.,
Moffat, A.~F.~J., \& Fitzgerald, M.~P.\ 1980, \aj, 85, 708

\bibitem[Feigelson et al.(2005)]{Feigelson05} Feigelson, E. D., et al. 2005, ApJS, 160, 379

\bibitem[Feigelson et al.(2007)]{2007prpl.conf..313F} Feigelson, E., 
Townsley, L., G{\"u}del, M., 
\& Stassun, K.\ 2007, in Protostars and Planets V, Eds., B. Reipurth, D. Jewitt, and K. Keil, University of Arizona Press, Tucson, 313 

\bibitem[Feigelson et al.(2011)]{F10} Feigelson, E.~D., et al.\ 2011, \apjs, submitted (CCCP Clustering Paper)

\bibitem[Flaccomio et 
al.(2003)]{2003A&A...402..277F} Flaccomio, E., Micela, G., \& Sciortino, S.\ 2003, \aap, 402, 277 

\bibitem[Gagn{\'e} et al.(2011)]{Gagne11} Gagn{\'e}, M., et al.\ 2011, \apjs, submitted (CCCP Massive Star Signatures Paper)

\bibitem[Getman et al.(2005)]{2005ApJS..160..319G} Getman, K.~V., et al.\ 
2005, \apjs, 160, 319 

\bibitem[Getman et~al.(2006)]{Getman06} Getman, K.~V., Feigelson,
  E.~D., Townsley, L., Broos, P., Garmire, G., \& Tsujimoto, M. 2006,
  \apjs, 163, 306

\bibitem[Getman et al.(2009)]{2009ApJ...699.1454G} Getman, K.~V., 
Feigelson, E.~D., Luhman, K.~L., Sicilia-Aguilar, A., Wang, J., 
\& Garmire, G.~P.\ 2009, \apj, 699, 1454 

\bibitem[Getman et al.(2010)]{G10} Getman, K.~V., Feigelson, E.~D.,
  Broos, P.~S., Townsley, L.~K., \& Garmire, G.~P.\ 2010, \apj, 708,
  1760

\bibitem[Getman et al.(2011)]{Getman11} Getman, K.~V., et al.\ 2011, \apjs, submitted (CCCP Contaminants Paper)  

\bibitem[Girardi et al.(2000)]{2000A&AS..141..371G} Girardi, L.,
  Bressan, A., Bertelli, G., \& Chiosi, C.\ 2000, \aaps, 141, 371

\bibitem[Grubissich(1968)]{G68} Grubissich, C.\ 1968,
Zeitschrift fur Astrophysik, 68, 173

\bibitem[Hamaguchi et al.(2009)]{2009ApJ...695L...4H} Hamaguchi, K., et 
al.\ 2009, \apjl, 695, L4 

\bibitem[Herbig(1960)]{1960ApJS....4..337H} Herbig, G.~H.\ 1960, \apjs, 4, 
337 

\bibitem[Hillenbrand \& Hartmann(1998)]{Hillenbrand98} Hillenbrand, L.~A., \& Hartmann, L.~W.\ 1998, \apj, 492, 540 

\bibitem[Hirschi et  al.(2004)]{Hirschi04} Hirschi, R., Meynet, G., \& Maeder, A.\ 2004, \aap, 425, 649 

\bibitem[Kissler-Patig et al.(2008)]{KP08} Kissler-Patig, M., et
  al.\ 2008, \aap, 491, 941

\bibitem[Kroupa(2001)]{2001MNRAS.322..231K} Kroupa, P.\ 2001, \mnras, 322, 
231 

\bibitem[Kumar et al.(2008)]{2008MNRAS.386.1380K} Kumar, B., Sagar, R., \& Melnick, J.\ 2008, \mnras, 386, 1380 

\bibitem[Lucy \& White(1980)]{Lucy80} Lucy, L.~B., \& White, R.~L.\ 1980, \apj, 241, 300 

\bibitem[Maschberger 
\& Kroupa(2009)]{2009MNRAS.395..931M} Maschberger, T., \& Kroupa, P.\ 2009, \mnras, 395, 931 

\bibitem[Mermilliod(1976)]{1976A&A....53..289M} Mermilliod, J.-C.\ 1976, \aap, 53, 289 

\bibitem[Meyer et al.(1997)]{1997AJ....114..288M} Meyer, M.~R., Calvet, N., 
\& Hillenbrand, L.~A.\ 1997, \aj, 114, 288 

\bibitem[Michael et al.(1983)]{Michael83} Michael J.R., 1983,
  Biometrika, 70, 1, pp.11-17

\bibitem[Morrell et al.(1988)]{M88} Morrell, N., Garcia,
B., \& Levato, H.\ 1988, \pasp, 100, 1431

\bibitem[Naz{\'e} et al.(2011)]{Naze11} Naz{\'e}, Y., et al.\ 2011, \apjs, submitted (CCCP Massive Star Lx/Lbol Paper)

\bibitem[Povich et al.(2011)]{Povich11} Povich, M.~S., et al.\ 2011, \apjs, submitted (CCCP IR YSOs Paper)

\bibitem[Preibisch et al.(2005)]{2005ApJS..160..401P} Preibisch, T., et 
al.\ 2005, \apjs, 160, 401 

\bibitem[Preibisch et al.(2011)]{P10} Preibisch, T., et al.\ 2011, \apjs, submitted (CCCP HAWK-I Paper)

\bibitem[Rachford 
\& Canterna(2000)]{2000AJ....119.1296R} Rachford, B.~L., \& Canterna, R.\ 2000, \aj, 119, 1296 

\bibitem[Sana et al.(2006)]{Sana06} Sana, H., Rauw, G., 
Naz{\'e}, Y., Gosset, E., \& Vreux, J.-M.\ 2006, \mnras, 372, 661 

\bibitem[Siess et al.(2000)]{Siess00} Siess, L., Dufour, E., \&
Forestini, M.\ 2000, \aap, 358, 593

\bibitem[Seward \& Chlebowski(1982)]{SC82} Seward, F.~D., \& Chlebowski, T.\ 1982, \apj, 256, 530 

\bibitem[Stelzer et al.(2005)]{2005ApJS..160..557S} Stelzer, B., Flaccomio, 
E., Montmerle, T., Micela, G., Sciortino, S., Favata, F., Preibisch, T., 
\& Feigelson, E.~D.\ 2005, \apjs, 160, 557 

\bibitem[Stelzer et~al.(2006)]{Stelzer06} {Stelzer}, B., {Micela}, G.,
  {Hamaguchi}, K., \& {Schmitt}, J.~H.~M.~M. 2006, \aap, 457, 223

\bibitem[Skiff(2009)]{Skiff09} Skiff, B.~A.\ 2009, VizieR
Online Data Catalog, 1, 2023

\bibitem[Smith et al.(2000)]{2000ApJ...532L.145S} Smith, N., Egan, M.~P., 
Carey, S., Price, S.~D., Morse, J.~A., 
\& Price, P.~A.\ 2000, \apjl, 532, L145 

\bibitem[Smith et al.(2001)]{2001ApJ...556L..91S} Smith, R.~K., Brickhouse, 
N.~S., Liedahl, D.~A., \& Raymond, J.~C.\ 2001, \apjl, 556, L91 

\bibitem[Smith(2006)]{S06} Smith, N.\ 2006, \mnras, 367, 763

\bibitem[Smith \& Brooks(2008)]{SB08} Smith, N., \& Brooks,
  K.~J.\ 2008, Handbook of Star Forming Regions, Volume II, 138

\bibitem[Tapia et al.(1988)]{T88} Tapia, M., Roth, M.,
Marraco, H., \& Ruiz, M.~T.\ 1988, \mnras, 232, 661

\bibitem[Tapia et al.(2003)]{T03} Tapia, M., Roth, M., 
V{\'a}zquez, R.~A., \& Feinstein, A.\ 2003, \mnras, 339, 44 

\bibitem[Townsley et al.(2011a)]{T11a} Townsley, L.~K., et al.\ 2011a, \apjs, submitted (CCCP Intro Paper)

\bibitem[Townsley et al.(2011b)]{Townsley11b} Townsley, L.~K., et al.\ 2011b, \apjs, submitted (CCCP Diffuse Paper)

%\bibitem[Townsley et al.(2011c)]{Townsley11c} Townsley, L.~K., et al.\ 2011, \apjs, submitted (CCCP Global Comparison Paper)

\bibitem[Th{\'e}
\& Vleeming(1971)]{T71} Th{\'e}, P.~S., \& Vleeming, G.\ 1971,
\aap, 14, 120

\bibitem[Th{\'e} et
al.(1980)]{T80} Th{\'e}, P.~S., Bakker, R., \& Antalova, A.\ 1980,
\aaps, 41, 93

\bibitem[Vazquez et 
al.(1996)]{1996A&AS..116...75V} Vazquez, R.~A., Baume, G., Feinstein, A., \& Prado, P.\ 1996, \aaps, 116, 75 

\bibitem[Vuong et al.(2003)]{2003A&A...408..581V} Vuong, M.~H., Montmerle, T., Grosso, N., Feigelson, E.~D., Verstraete, L., \& Ozawa, H.\ 2003, \aap, 408, 581 

\bibitem[Walborn(1973)]{W73} Walborn, N.~R.\ 1973, \apj, 
179, 517 

\bibitem[Walborn(1995)]{W95} Walborn, N.~R.\ 1995, Rev.\
Mex.\ Astro.\ Astrofis.\ Conf.\ Ser., 2, 51

\bibitem[Wang et al.(2007)]{Wang07} Wang, J., Townsley, L.~K.,
  Feigelson, E.~D., Getman, K.~V., Broos, P.~S., Garmire, G.~P., \&
  Tsujimoto, M.\ 2007, \apjs, 168, 100

\bibitem[Wang et al.(2008)]{Wang08} Wang, J., Townsley, L.~K.,
  Feigelson, E.~D., Broos, P.~S., Getman, K.~V.,
  Rom{\'a}n-Z{\'u}{\~n}iga, C.~G., \& Lada, E.\ 2008, \apj, 675, 464

\bibitem[Wang et al.(2010)]{Wang10} Wang, J., Feigelson, E.~D.,
  Townsley, L.~K., Broos, P.~S., Rom{\'a}n-Z{\'u}{\~n}iga, C.~G.,
  Lada, E., \& Garmire, G.\ 2010, \apj, 716, 474

\bibitem[Wilms et al.(2000)]{2000ApJ...542..914W} Wilms, J., Allen, A., 
\& McCray, R.\ 2000, \apj, 542, 914 

\bibitem[Winston et al.(2010)]{Winston10} Winston, E., et al.\ 
2010, \aj, 140, 266 

\bibitem[Wolk et al.(2010)]{Wolk10} Wolk, S.~J., Winston, E., 
Bourke, T.~L., Gutermuth, R., Megeath, S.~T., Spitzbart, B.~D., 
\& Osten, R.\ 2010, \apj, 715, 671 

\bibitem[Wolk et al.(2011)]{Wolk11} Wolk, S.~J., et al.\ 2011, \apjs, submitted (CCCP Tr16 Paper)

\end{thebibliography}
\end{document}